\documentstyle[prl,aps,multicol,mypsfig2]{revtex}

\def\be{\begin{equation}}
\def\ee{\end{equation}}
\def\bea{\begin{eqnarray}}
\def\eea{\end{eqnarray}}

\def\>{\rangle}
\def\<{\langle}

\newcommand{\mypsfig}[2]{\psfig{file=#1,#2}}

\def\<{\langle}
\def\>{\rangle}

\begin{document}

\title{Quantum teleportation is a universal computational primitive}

\author{
    Daniel Gottesman$^{1,2}$ \thanks{Electronic address:
                gottesma@microsoft.com}
    ~{\em and}~
    Isaac L. Chuang$^{3}$  \thanks{Electronic address:
                ichuang@almaden.ibm.com}
}

\address{\vspace*{1.2ex}
    $^1$ Theoretical Astrophysics T-6, MS B-288, \\[0.2ex] {}\vspace*{-0.8ex}
        Los Alamos National Laboratory, Los Alamos, NM 87545}
\address{\vspace*{-0.2ex}
    $^2$    Microsoft Research, One Microsoft Way 
        Redmond, WA 98052}
\address{\vspace*{-0.2ex}
    $^3$    IBM Almaden Research Center, 650 Harry Road 
        San Jose, CA 95120 }
\maketitle


\begin{abstract}
We present a method to create a variety of interesting gates by
teleporting quantum bits through special entangled states.  This
allows, for instance, the construction of a quantum computer based
on just single qubit operations, Bell measurements, and GHZ
states.  We also present straightforward constructions of a wide
variety of fault-tolerant quantum gates.
\end{abstract}

\begin{multicols}{2}

Creating a quantum computer capable of realizing the theoretical promise of
algorithms such as quantum factoring\cite{Shor94} and quantum
search\cite{Grover97} will require both a design for a large system capable of
very accurate controlled unitary evolution, and good fault-tolerant procedures
to overcome inevitable residual imperfections in the physical realization of
this system\cite{Preskill98b,Steane99a,Gottesman98a}.  There are many
suggested designs for quantum computers, but none are completely satisfactory,
in the sense that none allows a large quantum computer to be built in the near
future\cite{Preskill98a}; and some universal fault-tolerant protocols are
known, but they can be quite complicated, and frequently require many
operations to produce a specific desired
transformation\cite{Shor96b,KLZ,Preskill98b,Steane99a,Gottesman98a}.

Here, we address aspects of both problems, and show how a single technique --
a generalization of quantum teleportation\cite{Bennett93a} -- reduces resource
requirements for quantum computation and unifies known protocols for
fault-tolerant quantum computation.  We show, for instance, that a quantum
computer can be constructed using just single quantum bit (qubit) operations,
Bell-basis measurements, and Greenberger-Horne-Zeilinger states\cite{GHZ90}.
We also present straightforward constructions for a new, infinite class of
fault-tolerant quantum gates.  By making use of specific, pre-computable
entangled states, these techniques vividly illustrate how entanglement can be
a valuable resource for computation.

The heart of our discussion rests in the power of entangling measurements.
Measurement, in its guise as an interface between the quantum and classical
worlds, is generally considered to be an irreversible operation, destroying
quantum information and replacing it with classical information.  In certain
carefully designed cases, however, this need not be true.  For example,
quantum teleportation\cite{Bennett93a} uses measurement to transfer quantum
information from one place to another, and programmable quantum
gates\cite{Nielsen97c} can be used to probabilistically transform quantum
information by an arbitrary quantum operation.  Quantum error correction also
allows a large set of quantum operations, including measurement, to be
reversed.

In all these applications, quantum information is preserved only in a subspace
of the measured system.  By selecting our initial state to lie in this
preserved subspace, we can ensure, paradoxically, that the measurement tells
us {\em nothing} about the quantum data.  Still, the measurement can be very
useful --- once it has been done, the data is transformed in one of a variety
of ways, indexed by the random measurement outcome.  In the case of quantum
teleportation or quantum error correction, this fact is used to restore the
data to its initial state.  Here, in contrast, we shall use quantum
teleportation to transform data into a new state, corresponding to the action
of some quantum gate which would otherwise be difficult or impossible to
perform.

\section{Unitary transforms by teleportation}

\begin{figure}[htbp]
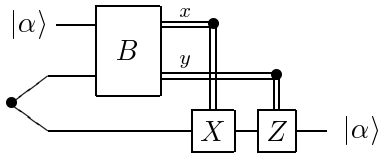

\begin{center}
\mbox{\mypsfig{Figure1.epsf}{scale=100}}
\end{center}
\caption{\narrowtext Quantum circuit for teleportation.  Time proceeds from
left to right.  $<$ denotes the EPR state $|\Psi\>$, and the box $B$ is a
measurement in the Bell basis.  The double wires carry classical bits, and the
single wires, qubits.  }
\label{fig:teleport}
\end{figure}

We begin by showing how a controlled-{\sc not} ({\sc cnot}) between two qubits
can be deterministically accomplished using quantum teleportation.  Recall how
quantum teleportation works: a single qubit state $|\alpha\> = a|0\> + b|1\>$
is prepared, along with an EPR state $|\Psi\> = (|00\>+|11\>)/\sqrt{2}$, then
$|\alpha\>$ and one qubit of $|\Psi\>$ are measured together in the Bell basis
$|0x\>+(-1)^z|1\bar{x}\>$ (where $x,z = \{0,1\}$, and $\bar{x} = 1-x$), giving
a (uniformly distributed) random two-bit classical result which is $xz$
\cite{Brassard96b}.  The output qubit is then in the initial state
$|\alpha\>$, but with an additional single-qubit Pauli operation $X$, $Y$, or
$Z$\cite{Gottesman-heisenberg} applied to it, with the random variable $xz$
determining which Pauli operator it is (with $00$ corresponding to the
identity).  We simply reverse the appropriate Pauli operator to reconstruct
$|\alpha\>$, as shown in Fig.~\ref{fig:teleport}.  Replication of this circuit
allows teleportation of multiple qubits.

\begin{figure}[htbp]
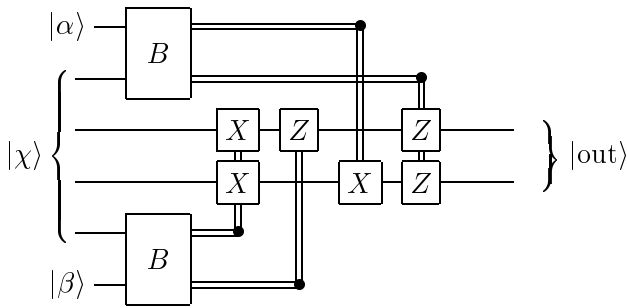

\begin{center}
\mbox{\mypsfig{Figure2.epsf}{scale=100}}
\end{center}
\caption{Quantum circuit for teleporting two qubits through a
    controlled-{\sc not} gate, giving $|{\rm out}\> = {\sc cnot}\,
    |\beta\>|\alpha\>$. }
\label{fig:dgate}
\end{figure}

The same basic idea can be used to teleport two qubits {\em
through a {\sc cnot} gate} (a two-qubit gate which flips
the ``target'' qubit whenever the ``control'' qubit is a $|1\>$);
that is, the reconstructed qubits are the original ones
transformed by a {\sc cnot} gate operation.  This is
accomplished by the circuit shown in Fig.~\ref{fig:dgate}, where
$|\alpha\> = a|0\> + b|1\>$ and $|\beta\> = c|0\> + d|1\>$ are two
arbitrary single qubit states, and \bea |\chi\> =
\frac{(|00\>+|11\>)|00\> + (|01\>+|10\>)|11\>}{\sqrt{2}} \,.  \eea
The {\sc cnot} gate has $|\beta\>$ as its control, and
$|\alpha\>$ as its target.

\begin{figure}[htbp]
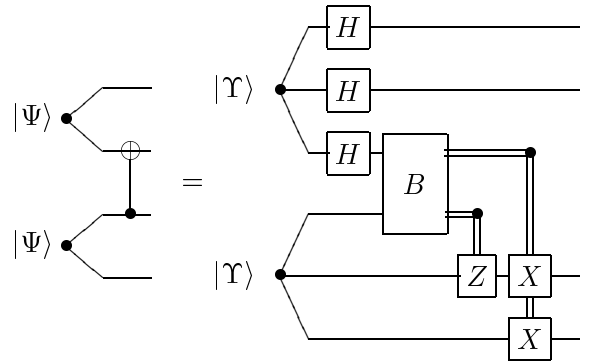

\begin{center}
\mbox{\mypsfig{Figure3.epsf}{scale=100}}
\end{center}
\caption{Quantum circuit to create the $|\chi\>$ state from two EPR pairs
(left), or from two GHZ states $|\Upsilon\> = (|000\>+|111\>)/\sqrt{2}$
(right). $H$ is the Hadamard gate.} \label{fig:chi}
\end{figure}

This can be verified by direct computation, but it is easier to understand by
realizing that $|\chi\>$ can be created simply using two EPR pairs
(Fig.~\ref{fig:chi}).  Combining this circuit with the previous one, we
immediately note that the only differences with Fig.~\ref{fig:teleport} are
the {\sc cnot} gate appearing between the two EPR pairs, and the
different classically controlled single qubit gates.  For each EPR pair, the
Bell basis measurement effectively introduces one of four random quantum
operations $(I,X,Y,Z)$ to the other half of the involved EPR pair, at a time
which is {\em before} the {\sc cnot} gate\cite{Nielsen97c}.

However, it happens that single Pauli operations occurring before
a {\sc cnot} gate are equivalent to (different) Pauli
operations occurring after the {\sc cnot}
gate\cite{Gottesman98a}.  For instance, ${\rm \sc cnot} (X \otimes
I) = (X \otimes X) {\rm \sc cnot}$.  This is equivalent to the
statement that conjugation by {\sc cnot} preserves the Pauli group
(comprised of tensor products of Pauli matrices, with overall sign
$\pm 1$).  Thus, the quantum teleportation construction still
works, but using different controlled single-qubit operations to
reconstruct the desired result.

This construction enables {\sc cnot} gates to be performed between
two qubits, using only classically controlled single qubit operations, prior
entanglement, and Bell basis measurements. Moreover, $|\chi\>$ can be created
from two pairs of GHZ\cite{GHZ90} states (Fig.~\ref{fig:chi}).

\section{Fault tolerant quantum computation}

Fault tolerant gates come from noting that essentially the same construction
works equally well for any gate $U$ which preserves the Pauli group under
conjugation; this set of gates, the Clifford group, plays an important role in
the theory of quantum error-correcting codes and
fault-tolerance\cite{Gottesman98a,Calderbank97a}.  To see how this is
accomplished, consider an $n$-qubit state $|\psi\>$, in which each qubit is
encoded using a stabilizer code, such as the $7$-qubit CSS
code\cite{Steane96c,Steane96a}.  $0$ and $1$ shall represent the corresponding
encoded qubit states.  Let $|\Psi^n\>$ be the $2n$-(encoded) qubit Bell state
$(|00\>+|11\>)^{\otimes n}$ (normalizations suppressed for clarity),
rearranged so that the first $n$ labels represent half of the EPR pairs (the
upper qubits), and the last $n$ the other half (the lower qubits).  In other
words, $(I\otimes U)|\Psi^n\>$ (where $I$ is the identity on $n$ qubits) is
$U$ acting on the lower qubits of all the EPR pairs.

The goal of fault-tolerant computation is to perform gates on the
logical qubits while restricting the propagation of errors among
the physical qubits, which can compromise the code's ability to
correct errors.  The usual method for doing this is to only perform
{\em transversal} gates on the code --- that is, gates which
interact qubits in one code block only with corresponding qubits
in other code blocks.  While errors may then propagate between
blocks, they cannot propagate within blocks, so a single faulty
gate can only cause a single error in any given block of the code.

Operators from the Pauli group (such as $X$, $Y$, and $Z$) can easily be
performed on logical qubits which are encoded with a stabilizer
code\cite{Gottesman97a}.  Let $C_1$ represent the Pauli group.  $C_2$, the
Clifford group, will be the set of gates which map Pauli operators into Pauli
operators under conjugation. Through an appropriate sequence of gates and
measurements, any $C_2$ operation can also be performed on any stabilizer
code\cite{Gottesman97a}.

More difficult to perform are gates in the class defined as
\be
    C_3 \equiv \{ U \,|\, U C_1 U^\dagger \subseteq C_2 \}
\,.
\ee
$C_3$ contains gates such as the Toffoli gate (controlled-controlled-{\sc
not}), the $\pi/8$ gate (rotation about the $Z$-axis by an angle $\pi/4$), and
the controlled-phase gate (${\rm diag}(1, 1, 1, i)$).  For instance, the
$\pi/8$ gate transforms $X \rightarrow PX $ and $Y \rightarrow -i PY$ ($Z$
commutes with the gate and is thus left unchanged), where $P$ is the $\pi/4$
gate (${\rm diag}(1, i)$).  Fault-tolerant constructions of these gates are
known\cite{Shor96b,KLZ,Roychowdhury99}, but they are {\it ad hoc} and do not
generalize easily.

\begin{figure}[htbp]
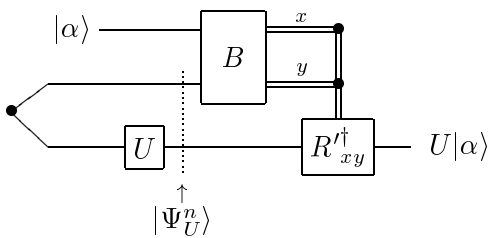

\begin{center}
\mbox{\mypsfig{Figure4.epsf}{scale=100}}
\end{center}
\caption{Quantum circuit to perform $U$ fault tolerantly using
quantum teleportation.  In general, this works for any $U\in C_k$,
since $R'_{xz}\in C_{k-1}$ by definition of $C_k$.}
\label{fig:ftqc-ck}
\end{figure}

However, our teleportation construction provides a straightforward
way to produce {\em any} gate in $C_3$, as shown in
Fig.~\ref{fig:ftqc-ck}.  For $U\in C_3$, first construct the state
\be
    |\Psi^n_U\> = (I\otimes U) |\Psi^n\>\,.
\ee
Next, take the input state $|\psi\>$ and do Bell
basis measurements on this and the $n$ upper qubits of
$|\Psi^n_U\>$, leaving us with $n$ qubits in the state
\be
    |\psi_{out}\> = U R_{xz} |\psi\> = R'_{xz} U |\psi\>
\,.
\ee
where $R_{xz}$ is an operator in $C_1$ which depends on the (random) Bell
basis measurement outcomes $xz$, and $R'_{xz}$ is an operator in $C_2$: the
image of $R_{xz}$ under conjugation by $U$.  Since $R'_{xz}$ is in the
Clifford group, it can be performed fault-tolerantly.  As long as
$|\Psi^n_U\>$ can be prepared fault-tolerantly, this construction allows $U$
to be performed fault-tolerantly.

Of course, $|\Psi^n_U\>$ must be prepared fault-tolerantly.  To do this, note
that the state $|\Psi^n\>$ is the $+1$ eigenvector of the $2n$ operators $X_i
\otimes X_i$ and $Z_i \otimes Z_i$ (where $X_i$ and $Z_i$ are $X$ and $Z$,
respectively, acting on the $i^{th}$ upper or lower qubit).  Therefore,
$|\Psi^n_U\>$ is the $+1$ eigenvector of the operators $M_i = X_i \otimes U
X_i U^\dagger $ and $ N_i = Z_i \otimes U Z_i U^\dagger $.  Furthermore, the
eigenvalues of these $2n$ operators completely determine the state, so if all
these operators have eigenvalue $+1$, the state actually is the desired one.
Therefore, to produce $|\Psi^n_U\>$, prepare $n$ EPR pairs, which can easily
be done fault-tolerantly by measuring $X_i \otimes X_i$ and $Z_i \otimes Z_i$
or with fault-tolerant Hadamard and {\sc cnot} gates, measure the operators
$M_i$ and $N_i$, and perform an appropriate Pauli operation $Z_i \otimes I$
or $X_i \otimes I$ as necessary to move into the $+1$ eigenspace of all the
$M_i$s and $N_i$s.

The hard part of the preparation is measuring $M_i$ and $N_i$
fault-tolerantly.  Since this construction is quite complicated in
the case where $M_i$ and $N_i$ do not have transversal
constructions, we defer the discussion of this point to the
Appendix.  Note, however, that the only point where this complex
construction is necessary is in the preparation of the ancilla
states $\Psi_U$ used in the teleportation.

\section{Conclusion}

Our construction of quantum gates using teleportation offers tantalizing
possibilities for relaxing experimental constraints on realizing quantum
computers.  For example, using single photons as qubits and current optical
technology, one can perform nearly perfect Bell basis
measurements\cite{Kwiat98a}, quantum teleportation\cite{Bouwmeester97a},
almost create GHZ states\cite{Zeilinger-ghz}, and certainly perform single
qubit operations\cite{Chuang95}.  Thus, given GHZ states, quantum computers
might be constructed nearly completely from linear optical components.
Similar implications can be drawn for other physical systems, particularly if
entangled states can readily be prepared and stored.

The construction of a fault tolerant Toffoli gate using teleportation is a
dramatic simplification of previous constructions, and generalizes through a
recursive application of the construction to provide an infinite family of
gates, $C_k \equiv \{ U | U C_1 U^{\dagger} \subseteq C_{k-1} \}$, all of
which can be performed fault tolerantly.  While the precise set of gates which
form $C_k$ is still under investigation, it is known that every $C_k$ contains
interesting gates, such as the $\pi/2^k$ rotations, which appear in Shor's
factoring algorithm\cite{Shor94}.  The states $|\Psi_U^n\>$ needed for a gate
in $C_k$ are exponentially difficult (in $k$) to construct, but they may be
prepared offline, since $|\Psi_U^n\>$ is independent of the data being acted
upon.  Thus, $|\Psi_U^n\>$ are valuable generic quantum resources which might
be considered a manufacturable commodity for quantum commerce!  Even if
$|\Psi_U^n\>$ are not available, the construction presents a great conceptual
simplification, and for small $k$, it can greatly reduce the number of
operations needed to assemble the precise gates called for in an algorithm,
clearly of benefit in efficiently performing quantum computation with
realistically imperfect gates.

\appendix
\section{Fault-Tolerant Preparation of $|\Psi_U^n\>$}

A crucial step in the fault-tolerant preparation of the ancilla state
$|\Psi_U^n\>$ is the measurement of an operator $M$ acting on the
logical qubits in a code.  The procedure for doing such measurements
has been described and is straightforward; however, its adaptation to
the present goal has not been clearly documented in the literature,
and there are a number of potential pitfalls which we believe are
useful to know about.

\begin{figure}[htbp]
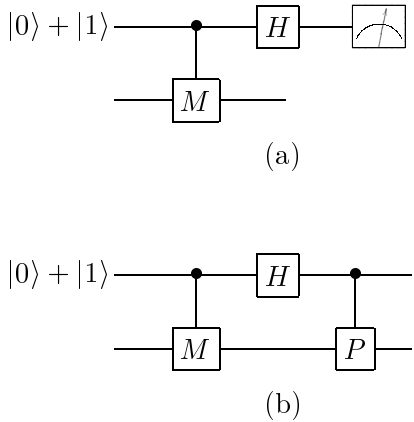

\begin{center}
\mbox{\mypsfig{Figure5.epsf}{scale=100}}
\end{center}
\caption{a) A non-fault-tolerant procedure to measure $M$ with
eigenvalues $\pm 1$, b) A coherent version of this procedure.}
\label{fig:measure}
\end{figure}

The basic problem can be illustrated by considering the standard
non-fault-tolerant measurement method shown in
Fig.~\ref{fig:measure}a. We prepare a control qubit is in the state
$|0\> + |1\>$, and perform a controlled-$M$ gate to the block of the
code containing the logical qubits we wish to measure.  The $M$ in
this case must be an encoded version of $M$, so it acts on the data,
and not the physical qubits making up the code.  Then we Hadamard
transform the control qubit and measure {\em it}.  It is sufficient
for our purposes to restrict attention to operators with eigenvalues
$\pm 1$; thus, the data will be collapsed on a $+1$ or $-1$ eigenstate
of $M$ when the control qubit reads $0$ or $1$, respectively.

This procedure fails to be fault-tolerant in a variety of ways.  The
control qubit is a single qubit; an error in it before or during the
operation could be propagated to every qubit in the code block. An
error on the control qubit after the measurement will tell us the
wrong value of the measurement, which may cause us to act improperly
later on.  Furthermore, in many cases of interest to us here, the
encoded version of $M$ is itself difficult to perform, requiring a
number of transversal operations and some measurements.  These
measurements, in turn, should be done fault-tolerantly, but since the
control qubits used for those measurements are entangled with the
control qubits one level up, they cannot simply be projectively
measured.

These problems can be solved using three basic ideas.  First, a
coherent measurement procedure can be adopted, utilizing
measurement results to immediately disentangle all involved
ancilla qubits. Second, by using multiple qubits prepared in a
``cat'' state $|00\cdots0\> + |11\cdots1\>$ of $n$ qubits (where
each block of the code also contains $n$ qubits), propagation of
errors can be limited.  Finally, since the state we are preparing,
$|\Psi_U^n\>$, is {\em known} beforehand (as opposed to computing
with variable data), performing the encoded version of $M$ can be
done through recursive application of the basic measurement
procedure.  These three steps are described in detail below.

Coherent measurement is possible in all cases of interest in this
paper.  In our application, the measurement, if it produces the
$-1$ eigenstate, is followed by a (classically controlled)
operation $P$ (frequently a Pauli operation) which moves the data
from a $-1$ eigenstate of $M$ to a $+1$ eigenstate of $M$.
Equivalently, we may instead follow the Hadamard transform on the
control qubit by a controlled-$P$ gate, as shown in
Fig.~\ref{fig:measure}b. Doing this leaves the control qubit
disentangled with the data (which is always in the same state, a
$+1$ eigenstate of $M$).  Of course, since this process still
depends on a single control qubit, it is not fault tolerant
either, so a modification of the control scheme is needed, using
cat states.

\begin{figure}[htbp]
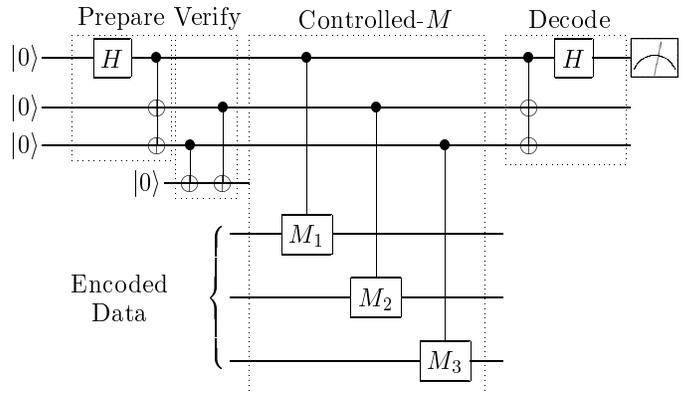

\begin{center}
\mbox{\mypsfig{Figure6.epsf}{scale=90}}
\end{center}
\caption{Fault-tolerant measurement of a gate $M$ with a
transversal implementation.} \label{fig:ftmeasure}
\end{figure}

Cat-state control is the method utilized in~\cite{Shor96b}
and~\cite{DiVincenzo96} to provide fault-tolerant measurement of
Pauli operators (elements of $C_1$).  As shown in
Fig.~\ref{fig:ftmeasure}, the single control qubit is replaced
with a cat state $|00\cdots0\> + |11\cdots1\>$ of $n$ qubits. Then
given a transversal implementation of $M$ (which, for a stabilizer
code, is always available for Pauli operators), we can easily
implement the controlled-$M$ part of the measurement: the gate of
$M$ which acts on the $k^{th}$ physical qubit of the block becomes
a controlled gate, conditioned on the $k^{th}$ qubit of the cat
state. Since in the absence of errors, every qubit in the cat
state is either $0$ or $1$, we either perform $M$ completely or
not at all.  If a single qubit of the cat state is wrong, the
error can only propagate to the corresponding qubit of the code.
The preparation of the cat state (which involves {\sc cnot}s
between the qubits of the state) might have resulted in multiple
errors, so before interacting it with the data, we should verify
the cat state by comparing pairs of qubits --- all should be the
same.

Afterwards, we decode the cat state with a series of {\sc cnot}
gates and a Hadamard; the resulting bit is again $0$ or $1$
depending on the eigenvalue of the state of the data. The value of
this result does still depend on the bottleneck of the single
qubit produced by decoding the cat state. In fact, even a single
phase error on one qubit of the cat state at any time will give us
the wrong value for the decoded cat state. Therefore, in order to
gain sufficient confidence in the result, we repeat the procedure
a number of times, and only act on the majority result. Also note
that a single error in the {\em data} might cause a wrong
measurement result, so we should perform error correction between
measurement trials.

The case of interest here is when $M$ is generally some element of
$C_k$ (not just $C_1$), in which case recursive application of the
above procedures is necessary, since implementation of $M$ will
generally consist of a series of transversal gates and measurements.
For instance, to prepare the ancilla needed to teleport a $C_3$ gate
$U$, we must measure gates of the form $M = P \otimes U P U^\dagger$,
where $P$ is some Pauli operator.  By the definition of $C_3$, $M$ is
in the Clifford group, and for a general stabilizer code, there will
be no simple transversal implementation of $M$.  The transversal gates
in the implementation of $M$ present no particular problem --- we can
condition them on the cat state just as in Fig.~\ref{fig:ftmeasure}.
The measurements present the difficulty.  Each will require its own
cat state (a sequence of them, in fact, since the measurement requires
a number of trials), and we must be certain the ``inner'' measurement
(of some operator $N$) does not destroy the superposition of the cat
state for the ``outer'' measurement of $M$.

We proceed as follows:  For each trial for the inner measurement,
prepare and verify a single cat state of $n$ qubits.  Using this
cat state, perform a controlled-$N$ operation in the usual way, as
per Fig.~\ref{fig:ftmeasure} (assume for the moment we have a
transversal implementation of $N$).  However, the gates making up
this operation are themselves controlled by qubits from the outer
cat state.  For instance, if $N$ requires {\sc not} gates on each
qubit in the code, we perform controlled-controlled-{\sc not}
gates instead, using the $k$th qubits of the inner and outer cat
states as controls for the $k$th qubit in the block.  Then we
decode the cat state normally.  If there have been no errors, the
result is a single qubit in the state $|0\>$ for the 0 part of the
outer cat state.  For the $1$ part of outer cat state, the qubit is
entangled with the data: $\alpha |0\> |\phi_0\> + \beta |1\>
|\phi_1\>$ (where $|\phi_0\>$ and $|\phi_1\>$ are eigenstates with
eigenvalues $\pm 1$, and the data begins in the state $\alpha
|\phi_0\> + \beta |\phi_1\>$). Since the ``data'' here is actually
an ancilla we are preparing, we know the values of $\alpha$ and
$\beta$.  This fact will be important later.

We repeat the above cat state preparation, controlled-$N$, and
decoding for each of the $n$ cat states in a trial.  Still
assuming no errors, the overall state of the system at this point
is
\begin{eqnarray}
& |00\cdots0\>_{\rm oc} |0\>_{\rm ic} |\phi\>_{\rm data} +
\label{eq:after} \\
& |11\cdots1\>_{\rm oc} \left(\alpha |0\>_{\rm
ic} |\phi_0\>_{\rm data} + \beta |1\>_{\rm ic} |\phi_1\>_{\rm
data}\right). \nonumber
\end{eqnarray}
The subscript ``oc'' indicates the outer cat states, the subscript
``ic'' indicates the qubit produced after decoding the inner cat
state for a single measurement trial, and $|\phi\> = \alpha
|\phi_0\> + \beta |\phi_1\>$ is the state of the data before the
measurement.

An error anywhere in this procedure could potentially give us the
wrong value for the decoded inner cat state, which is why we need
more measurement trials.  We should check, however, that a single
error in the procedure will only cause a single error in the data.
This is true, in fact: an error in a single qubit of the inner or
outer cat states can only propagate to the corresponding qubit in
the data block.

To prevent a repetition of any problem that may have occurred in
the first trial, we perform an error correction operation on the
code block, and reverify the outer cat state, correcting any
mistakes we see in that state. Then we go through another complete
trial. We continue alternating measurement trials with error
correction/verification steps for a total of $r$ trials (for some
$r$ large enough to give us confidence in the result).  Assuming
no errors, the result will look like Eq.~(\ref{eq:after}), except
there will now be a total of $r$ inner qubits, which in the
absence of errors would all be the same. For each of the $n$
qubits in the data block, we take the majority value of the $r$
inner qubits and store the result in a new ancilla qubit.  Then
we perform the controlled-$P$ ``correction'' step (as per
Fig.~\ref{fig:measure}b) based on the majority value for {\em
just} the single qubit at that coordinate. Therefore, a single
error in the majority calculation will only affect a single qubit
in the data block.

The final step in the recursive construction is to appropriately
disentangle the inner and outer qubits.  Assuming no errors, when the
outer cat state is $0$, the inner qubits are all in the state $|0\>$
as well, and the data block is in the state $|\phi\>$. When the outer
cat state is $1$, the data block is in the state $|\phi_0\>$ (as
desired), but the $r + n$ inner qubits are in a superposition $\alpha
|00\cdots0\> + \beta |11\cdots1\>$, so the inner qubits are still
entangled with the outer cat state.  Therefore, we perform sufficient
{\sc cnot} operations among the inner qubits to leave one in the state
$\alpha |0\> + \beta |1\>$, and the others all as $|0\>$.  Now we use
our knowledge of $\alpha$ and $\beta$ to rotate the single remaining
inner qubit back to $|0\>$, conditioned on (any) single qubit from the
outer cat state.  This completely disentangles the inner qubits from
the outer cat state and the data, leaving us with the state
\begin{equation}
|00\cdots0\>_{\rm oc} |\phi\>_{\rm data} + |11\cdots1\>_{\rm oc}
|\phi_0\>_{\rm data},
\end{equation}
as desired.

As we noted before, a single error during any trial propagates to
at most one qubit in the data block.  A single qubit error in the
outer cat state or the data block will ruin an inner measurement
trial, but will not survive the subsequent verification and error
correction step, so it only ruins the one set.  Since we perform
$r$ inner measurement trials, a total of $r/2$ such errors will be
required to ruin every majority calculation.  Therefore, for large
enough $r$, this will be of the same order of magnitude as other
failure modes (such as having many errors in the data block
itself). An error in a single majority calculation will only
produce a single error in the data block.  There are a number of
places, however, where a single error can cause the
disentanglement of the inner qubits to fail.  This will
effectively collapse the superposition of $0$ and $1$ in the outer cat
state. This is annoying, but not fatal; a single qubit error
directly in the outer cat state can produce the same result, which
is one reason we require a number of trials for any measurement.

We have demonstrated a procedure which performs a inner
measurement conditioned on an outer cat state.  By stringing these
together with transversal operations, we can measure any operator
$M$ in the Clifford group.  This allows us to create ancillas to
teleport any $C_3$ gate.  For $C_4$ and higher gates, we will need
similar, but more complicated procedures.  We will need to measure
$C_3$ gates; this requires the production of an ancilla for the
$C_3$ gate, which in turn requires measurement of a $C_2$
(Clifford group) gate, which may require measurement of Pauli
group operators.  Therefore, we may require 3 levels of cat states
at any given time, but by simply nesting the above procedure, we
can also produce the ancillas needed for $C_4$ gates.  Further
nesting will allow us to build gates from $C_5$ and higher.



\begin{thebibliography}{10}

\bibitem{Shor94} Shor, P., `Algorithms for quantum computation: discrete
logarithms and factoring,' {\em Proc. 35$^{th}$ Ann. Symp. on
Found. of Computer Science} (IEEE Comp. Soc. Press, Los Alamitos,
CA, 1994) 124--134, quant-ph/9508027.

\bibitem{Grover97} Grover, L.~K., `Quantum computers can search arbitrarily
large databases by a single query,' Phys. Rev. Lett. {\bf 79}, 23, 4709--4012
(1997).

\bibitem{Preskill98b}
Preskill, J., `Reliable quantum computers,' Proc. Roy. Soc. A:
Math., Phys. and Eng. {\bf 454}, 385--410  (1998),
quant-ph/9705031.

\bibitem{Steane99a}
Steane, A.M., `Efficient fault tolerant quantum computing,' Nature
{\bf 399} 124--6 (1999), quant-ph/9809054.

\bibitem{Gottesman98a}
Gottesman, D., `Theory of fault-tolerant quantum computation,'
Phys. Rev. A {\bf 57},  127--137 (1998), quant-ph/9702029.

\bibitem{Preskill98a}
Preskill, J., `Quantum computing: pro and con,' Proc. R. Soc.
Lond. A, {\bf 454}, 469--86 (1998), quant-ph/9705032.

\bibitem{Shor96b}
Shor, P.~W., `Fault-tolerant quantum computation,' {\em
Proceedings, 35th Annual Symposium on Fundamentals of Computer
Science} (IEEE Press, Los Alamitos, 1996) 56--65,
quant-ph/9605011.

\bibitem{KLZ}
Knill, E., Laflamme, R., and Zurek, W., `Resilient quantum computation,'
Science {\bf 279}, 342--345 (1998).

\bibitem{Bennett93a}
Bennett, C.~H.  {\it et~al.}, `Teleporting an Unknown Quantum State via Dual
Classical and {EPR} Channels,' Phys. Rev. Lett. {\bf 70},  1895--1899
(1993).

\bibitem{GHZ90}
Greenberger, D., Horne, M., Shimony A., and Zeilinger, A.,
`Bell's theorem without inequalities,' Amer.\ J.\
Phys. {\bf 58},  1131--43 (1990).

\bibitem{Nielsen97c}
Nielsen, M.~A.  and Chuang, I.~L., `Programmable quantum gate
arrays,' Phys. Rev. Lett. {\bf 79},  321--324 (1997),
quant-ph/9703032.

\bibitem{Brassard96b}
Brassard, G.,  `Teleportation as a quantum computation,'
{\em PhysComp 96}, edited by T. Toffoli, M. Biafore, and J.
  Leao (New England Complex Systems Institute, Cambridge MA, 1996), pp.\
  48--50, quant-ph/9605035.

\bibitem{Gottesman-heisenberg}
Gottesman D., "The Heisenberg Representation of Quantum
Computers," in {\it Group22: Proceedings of the XXII International
Colloquium on Group Theoretical Methods in Physics}, eds. S. P.
Corney, R. Delbourgo, and P. D. Jarvis, pp. 32-43 (Cambridge, MA,
International Press, 1999); LANL E-print quant-ph/9807006.

\bibitem{Calderbank97a}
Calderbank, A.~R., Rains, E.~M., Shor, P.~W., and Sloane,
N.~J.~A., `Quantum error correction and orthogonal geometry,'
Phys. Rev. Lett. {\bf 78},  405--8  (1997), quant-ph/9605005.

\bibitem{Steane96c} Steane A.~M., `Multiple particle interference and quantum
error correction,' Proc. Roy. Soc. Lond. A {\bf 452}, 2551--76
(1996), quant-ph/9601029.

\bibitem{Steane96a}
Steane, A.~M., `Error correcting codes in quantum theory,' Phys.
Rev. Lett.{\bf 77},  793--7  (1996).

\bibitem{Gottesman97a}
Gottesman, D., `Stabilizer codes and quantum error correction,'
Ph.D. thesis, California Institute of Technology, Pasadena, CA,
1997, quant-ph/9705052.

\bibitem{Roychowdhury99}
Boykin, P.~O., Mor, T., Pulver, M., Roychowdhury, V., and
F.~Vatan, "On Universal and Fault-Tolerant Quantum Computing,"
LANL E-print quant-ph/9906054.

\bibitem{Chuang95}
Chuang, I.~L. and Yamamoto, Y., `Simple Quantum Computer,' Phys.
Rev. A {\bf 52},  3489--3496 (1995), quant-ph/9505011.

\bibitem{Kwiat98a} Kwiat, P.~G. and Weinfurter, H., `Embedded Bell-state
analysis,' Physical Review A {\bf 58}, R2623--6 (1998).

\bibitem{Bouwmeester97a}
Bouwmeester D.  {\it et~al.}, `Experimental quantum teleportation,'
Nature {\bf 390},  575--9  (1997).

\bibitem{Zeilinger-ghz}
Bouwmeester, D.  {\it et~al.}, `Observation of three-photon
Greenberger-Horne-Zeilinger entanglement,'
Phys. Rev. Lett., {\bf 82},
1345--9 (1999).

\bibitem{DiVincenzo96}
DiVincenzo, D.~P.~and Shor, P.~W., `Fault-Tolerant Error
Correction with Efficient Quantum Codes,' Phys. Rev. Lett. {\bf
77}, 3260 (1996), quant-ph/9605031.

\end{thebibliography}


\end{multicols}

\end{document}